\documentstyle[emulateapj,psfig]{article}
\def\thesaurus#1{}
\def\and{and}
\def\offprints#1{}
\def\inst#1{}
\begin{document}
%
\title{A method for optimal image subtraction.} 
\author{C. Alard \altaffilmark{1,2}, R.~H. Lupton \altaffilmark{3}}
\altaffiltext{1}{DASGAL, 61 Avenue de l'observatoire, F-75014
 Paris, France}
\altaffiltext{2}{Institut d'Astrophysique de Paris, 98bis Boulevard Arago, F-75014 Paris, France}
 \altaffiltext{3}{Peyton Hall, Princeton}
 \offprints{C. Alard}
\date{Received ......; accepted ......}
\begin{abstract}
We present a new method designed for optimal subtraction of two
images with different seeing. 
Using image subtraction appears to be essential for the full
analysis of the microlensing survey images, however a perfect
subtraction
of two images is not easy as it requires the derivation of an
extremely accurate convolution kernel. Some empirical attempts to find the
kernel have used the Fourier transform of bright stars,
but solving the statistical problem of finding the best kernel solution has
never really been tackled. 
We demonstrate that it is possible to derive an optimal kernel solution
from a simple least square analysis using all the pixels of both images, and
also show that it is possible to fit the differential background variation
at the same time. We also show that PSF variations can also be easily handled
by the method.
 To demonstrate the practical efficiency of the method, we analyzed some
images from a Galactic Bulge field monitored
 by the OGLE II project.
We find that the residuals in the subtracted images
 are very close to the photon noise expectations. We also present some
light curves of variable stars, and 
 show that, despite high crowding levels, we get an error distribution
close to that expected from photon noise alone.
We thus demonstrate that nearly optimal differential photometry
can be achieved even in very crowded
 fields. We suggest that this algorithm might be particularly important
for microlensing surveys, where
 the photometric accuracy and completeness levels could be very
significantly improved by using this method.
\end{abstract}
\section{INTRODUCTION}
The search for microlensing events towards the LMC MACHO (Alcock {\it et al.}  1993), EROS (Aubourg {\it et al.} 1993)
the Galactic Bulge OGLE (Udalski {\it et al.} 1994) , MACHO, DUO  
(Alard \& Guibert 1997) or the M31 Galaxy, AGAPE (Ansari {\it et al.} 1997),
has provided us with an impressive database of images of densely crowded 
fields. The target fields have been monitored for several seasons, providing us
with
 time series containing hundreds of
images.                                      Light curves for millions
of stars can then be easily obtained with one of the widely
 used profile fitting codes such as DoPHOT (Schechter \& Mateo, 1993).
The search
 for variable objects among these huge light curve databases has proved
 very fruitful, for microlensing (MACHO, OGLE, DUO, EROS), and also for
variable stars (MACHO, OGLE, DUO, EROS). However we would like to emphasize
that photometry
and detection of variable (including moving) objects 
 should be based on the difference between frames, whereas
photometric codes like DoPHOT are designed to perform profile fitting
photometry of stars detected on a reference frame. 
In a variable object appears but was not seen on the reference it
won't be detected, leading
to a serious loss of efficiency for microlensing. The completeness of the 
variable star catalogue will be also
seriously affected. Another concern is that of photometric accuracy. With
multi-profile fitting techniques, the absolute photometry
 of a given (crowded) star requires perfect PSF estimation and careful
modeling of all
other close components, and also a correct estimate of the background value 
around each star. For the particular application of finding light curves
of variable objects, it is more efficient to estimate only that part of
the star's brightness which varies from image to image; this is exactly
the problem that image subtraction is designed to solve.
 The first attempt to perform image subtraction was made by Tomaney 
 \& Crotts 1996, (hereafter TC) for data taken towards the M31 Galaxy (Crotts \& Tomaney 1996). 
 To make a perfect subtraction of two images, one has 
 to match the frames to exactly the same seeing. TC proposed 
degrading a good seeing image to match a reference frame with bad seeing.
The quality achieved in the subtracted image is very dependent
of the quality of the kernel determination, and finding the proper kernel
is a very delicate operation. TC proposed
deriving the kernel by simply taking the ratio of the Fourier transform of a
bright star on each image. However the high frequencies are dominated
by noise and they were forced to use a Gaussian extrapolation to determinate
 the
wings (Phillips and Davies 1995). This method provides no guarantee
of producing the highest attainable quality of the subtracted image. Even apart
from the non-Gaussian wings of the true kernel, and the limited number of
bright, uncrowded, stars with sufficient signal to noise ratio, this 
method is non-optimal in the sense that it
does not use all the information available: in fact every star, even if
extremely crowded, contains information about the kernel; to get an optimal
solution we must use all of that information.
Additionally their method
has difficulty with rapid, complicated, PSF variations, and
does not intrinsically handle background subtraction.
The problem that we address here is how to find
an optimal kernel solution, in order to get the best possible subtracted
 image.
\section{The method.}
\subsection{Preliminaries.}
Before looking for the optimal subtraction, we need to perform some basic
 operations, to register the frames to a reference frame. 
 Usually the frames
 have slightly different centers and orientation (and possibly scale), and
 we need to perform an astrometric transform to match the coordinates of the
 reference frame. We determine this transform 
 by fitting a two dimensional polynomial using 500 stars on the reference
 frame, and the same number on the other frame. Using this transform we then
 resample the
 frame on the grid defined by the reference frame. This resampling is performed
 by interpolating using bicubic splines, which gives excellent accuracy. All
 the frames are then on the same coordinate system, and we can proceed
 to matching the seeing.
\subsection{The reference frame.}
Here we emphasize
 that, contrary to TC, we choose to take the {\it best}
 seeing frame as the reference. We do not wish to degrade the frame to
 the worst seeing
 frame, as this will clearly lower the signal to noise ratio. Later,
 we will match the seeing in our frames by convolving the reference to
 the seeing of each other frame. This is likely to be more difficult,
 as aligning to good seeing frames is more difficult, but we are looking
 for an optimal result.
\subsection{Seeing alignment to the reference.}
\label{matchingSeeing}
We now arrive to the fundamental problem of matching the seeing of two frames
 with different PSFs. We do not want to make any assumption concerning the
 PSF on the frame, and we plan to use all the pixels. The important point is
 that most of the stars on a given frame do not have large amplitude 
variations, but variations
 of at most 1 or 2 \%. This allows us to say that most of the pixels on two
 frames of the same field would be very similar, if the seeing were the same.
Consequently, one possibility is to try to find
the kernel by finding the least square solution of the equation:
\begin{equation}
 \hbox{Ref}(x,y) \otimes \hbox{Kernel}(u,v) = I(x,y)
\end{equation}
Where Ref is the reference image, and I the image to align. The symbol
 $\otimes$ denotes convolution.
In principle solving this equation is a non-linear problem, for which a 
realistic computer solution looks impossible. However if we decompose 
our kernel using some basis of functions, the problem becomes
a standard linear least square problem. If we decompose the kernel as:
\begin{equation}
 \hbox{Kernel}(u,v) = \sum_i \ a_i \times B_i(u,v),
\end{equation}
solving the least square gives the following Matrix equation for the $a_i$
coefficients:
$$
 M {\bf a} = {\bf V}
$$
With:
$$
 M_{ij} = \int C_i(x,y) \times C_j(x,y)/\sigma(x,y)^2 \ dx\, dy
$$
$$
 V_i = \int \hbox{Ref}(x,y) \times C_i(x,y)/\sigma(x,y)^2 \ dx\, dy
$$
$$
 C_i(x,y) = I(x,y) \otimes B_i(x,y)
$$
In choosing to solve the problem by least-squares we've implicitly
approximated the images Poisson statistics with Gaussian distributions with
 variance $\sigma(x,y)^2$:
$$
 \sigma(x,y) = k \times \sqrt{I(x,y)}
$$
 We set the constant k by taking into account the detector's gain
(ratio of photons detected to ADU). Note that the matrix M is just the
scalar product of the set of vectors $C_i$, and the vector V the scalar
product of the $C_i$ with I.
All we have to do now is to look for a suitable basis of functions to 
model the kernel. The functions of this basis must have finite
sums, and must drop rapidly beyond a given distance (the size of an isolated
star's image). To solve this problem, we start with a set of
Gaussian functions, which we modify by multiplying with a polynomial.
These basis functions allow us to model the kernel, even
if its shape is extremely complicated. We adopt the following decomposition:
$$
\hbox{Kernel}(u,v) = \sum_n  \sum_{d^x_n} \sum_{d^y_n}\, a_n \times
	e^{-(u^2 + v^2)/2\sigma_n^2} \,  u^{d^x_n}\, v^{d^y_n}
$$
where $0 < d^x_n \le D_n$, $0 < d^y_n + d^x_n \le D_n$, and $D_n$ is the degree
of the polynomial corresponding to the ${\rm n^{th}}$ Gaussian component.
There are a total of $(D_n + 1)(D_n + 2)/2$ terms for each value of $n$.

In the notation of eq (2),
$$
B(u,v) \equiv e^{-(u^2 + v^2)/2\sigma_n^2} \, u^{d^x_n}\, v^{d^y_n}
$$
In practice, it seems that 3 Gaussian components with associated polynomial
degrees
in the range 2 to 6 can give subtracted images with residuals comparable to 
$\sqrt{2}\times\hbox{photon noise}$.
\subsection{Differential background subtraction.}
Another important issue is that the differential background variation between
 the frames can be fitted simultaneously with the kernel. In eq (1) we did not
 considered any background variations between the two frames; let's modify eq
 (1)
in the following way:
\begin{equation}
  \hbox{Ref}(x,y) \otimes \hbox{Kernel}(x,y) = I(x,y) + bg(x,y)
\end{equation} 
We shall use the following polynomial expression for $bg(x,y)$:
$$
  bg(x,y) = \sum_i \sum_j a_i x^i \, y^j
$$
with $0 < i \le D^{bg}$, $0 < i + j \le D^{bg}$, and $D^{bg}$ is the
degree of the polynomial used to model the differential background variation.
The least square solution of eq (3), will lead to a matrix equation similar to
the previous one, except that we have to increase the number of $C_i$ vectors;
our definitions of the matrix M and vector V relative to the $C_i$ remain the
same as in section \ref{matchingSeeing}. We have:
$$
C_i(x,y) = \left\{\begin{array}{l@{\quad\hbox{if}\quad}l}
			x^j \, y^k, &
				i=0 \cdots n_{bg}-1 \\
			I(x,y) \otimes B_i(x,y), &
				i = n_{bg} \cdots n_{bg}+n \\
		\end{array}\right.
$$
where $n_{bg}=(D^{bg}+1)(D^{bg}+2)/2-1$ and $n = \sum_j (D_j+1) (D_j+2)/2$;
note that, for $i >= n_{bg}$, the $C_i$ are identical to our previous results.
\section{Taking into account the PSF variations.}
There are two ways to handle the problem of PSF variations.
Firstly, most of the time, the field is so dense that a
transformation kernel can be determined in small areas, small enough that
we can ignore the PSF's variation. This is the great advantage of a
method which does not require any bright isolated stars to determine
the kernel, but can be used on any portion of an image, provided that
the signal to noise is large enough to determine the kernel.  Indeed,
the more crowded the field the easier it is to model variations of the
PSF.
A second possibility is to make an analytical model of the kernel 
variations. We take the following kernel model:
\begin{eqnarray*}
\hbox{Kernel}(x,y,u,v) &=&
      \sum_n \sum_{d^x_n} \sum_{d^y_n} \sum_{\delta^x} \sum_{\delta^y} 
	\left[
		a_n \, x^{\delta^x} \, y^{\delta^y} \right.\\
& & \qquad 
\times \left.	e^{-(u^2 + v^2)/2\sigma_n^2} \, u^{d^x_n} \, v^{d^y_n}
	\right] \\
\end{eqnarray*}
Where:
$0 < \delta^x < D^k$, $0 < \delta^y  + \delta^x \le D^k$,
and $D^k$ is the degree of
the polynomial transform that we use to fit the kernel variations.
Provided that the kernel variations with x and y are small enough compared
to the u,v variations, we can easily calculate new expressions for the $C_i$s:
$$
C_i(x,y) = \left\{\begin{array}{l@{\quad\hbox{if}\quad}l}
			x^j \, y^k, &
				i=0 \cdots n_{bg}-1 \\
			I(x,y) \otimes Bo_i(x,y), &
				i = n_{bg} \cdots n_{bg}+n \\
		\end{array}\right.
$$
where
$$
Bo_i(u,v) \equiv B_i(u,v) \times u^{\delta^x} \, v^{\delta^y}
$$
with the values of $\delta_x$ and $\delta_y$ implicit in the index $i$,
and now $n = \sum_j (D_j+1) (D_j+2)/2) \times (D^k+1)(D^k+2)/2$.
Unfortunately these equations do not guarantee the conservation of flux.
 Consequently we must add the condition that the sum of the kernel has to
 be constant. To simplify the equation we also normalize the Bo functions,
 so that each of them sums to one. We can then rewrite the kernel decomposition:
\begin{eqnarray*}
 \hbox{Kernel}(x,y,u,v) &=& \sum_{i=0}^{n-1} a_i(x,y) \times
                                                \left [
Bo_i(u,v)-Bn_i(u,v)\right] \\
& & \qquad + \ \hbox{norm} \times Bo_n
\end{eqnarray*}
We can calculate the norm (the sum of the kernel) by making a constant PSF fit
in several small area. The different values will then be averaged to get the
constant norm.
The solution of the system for the coefficients $a_n$ is very similar
to the previous case of a constant PSF. We shall not bother to give all the
the details here.
\section{Application of the method to OGLE data.}
The OGLE team has kindly provided us with a stack of images of a field situated
 2 degrees from the Galactic Center, in order to experiment with our method. 
 For these particular images, the optimal kernel has a complicated shape and
it would be probably be very
 difficult to compute reliably with a simple Fourier division; we consider this
field an excellent test of our method. The data was taken in drift scan mode
(TDI), so the form of the PSF can vary rapidly with row number on the CCD.
We extracted
a small ($500\times1000$) sub-frame from the $2048 \times 8192$ original
images. One of the images has
quite outstanding seeing, and we took it as a reference. All frames were
 resampled to the reference grid by using the method previously described.
 To model the kernel, we took 3 Gaussian components with associated
 polynomials. For the first
 Gaussian we took $\sigma=1$ pixels and $\sigma=3$ and $\sigma=9$ for the two
 others. The degree
 of the associated polynomials were respectively 6,~4,~and~2. We divided
 the sub frame into $128 \times 256$
 pixels regions. We applied our method to each of these regions, which
 provided us with one subtracted
 image per region. We reconstructed the subtracted image of our whole
 sub-frame by mosaicing the subtracted
 images obtained for each region. In this set of 86 images, the seeing
 varies from 0.7 arcsec to 2.5 arcsec, and some of the frames have
 elongated stellar images.
We started by making an initial residual image using all unsaturated pixels.
We then made a 3 $\sigma$ rejection of the pixel list, to get rid of the
variables. We usually used 4 iterations of the method, to be completely
unbiased by large amplitude variables.
We found that for all images, the final residual calculated from 
the subtracted image was very close to that expected from Poisson statistics.
To illustrate this result we plot in Fig. 1 the initial images and the
 subtracted image for a small field containing a variable star at its center.
The stellar images are sharply peaked on the reference, while they look quite
 fuzzy and assymetric on the other image. This is well confirmed by the shape
 of the best convolution kernel which looks elongated and has a complicated 
shape. This example clearly illustrate the ability of our method to deal with
 any kernel shape. We can imagine that in this case any Gaussian 
 approximation of the kernel itself or of its Fourier transform would not 
 be satisfactory. For illustrative purposes, we
 also normalized the subtracted image by the sum of the photon noise expected from 
 the two images (see Fig. 2). Once 
 this normalization is applied, we see that the larger deviations visible at
 the location of the bright stars disappear, suggesting that the subtraction
 errors correspond to Poisson noise. This is confirmed by calculating the
 reduced chi squared:
 we find $\chi^2/\nu = 1.05$ (before doing this calculation we removed a small
 area around the variable star at center of the image). We also plot the
 histogram of the normalized deviations in Fig. 2. This histogram is very
 close to a 
 Gaussian with zero mean and unit variance (i.e. N(0,1)).
We observe deviations significantly larger than the Poisson
 expectations only for very bright stars (about 5 to 10 times brighter than
 the brightest stars in the small field we present).
We believe that these residuals are due
 to seeing variations, see section 5 for more details; the number
 of such bright stars in an image is very small.
\begin{figure*}
\centerline{\psfig{angle=0,figure=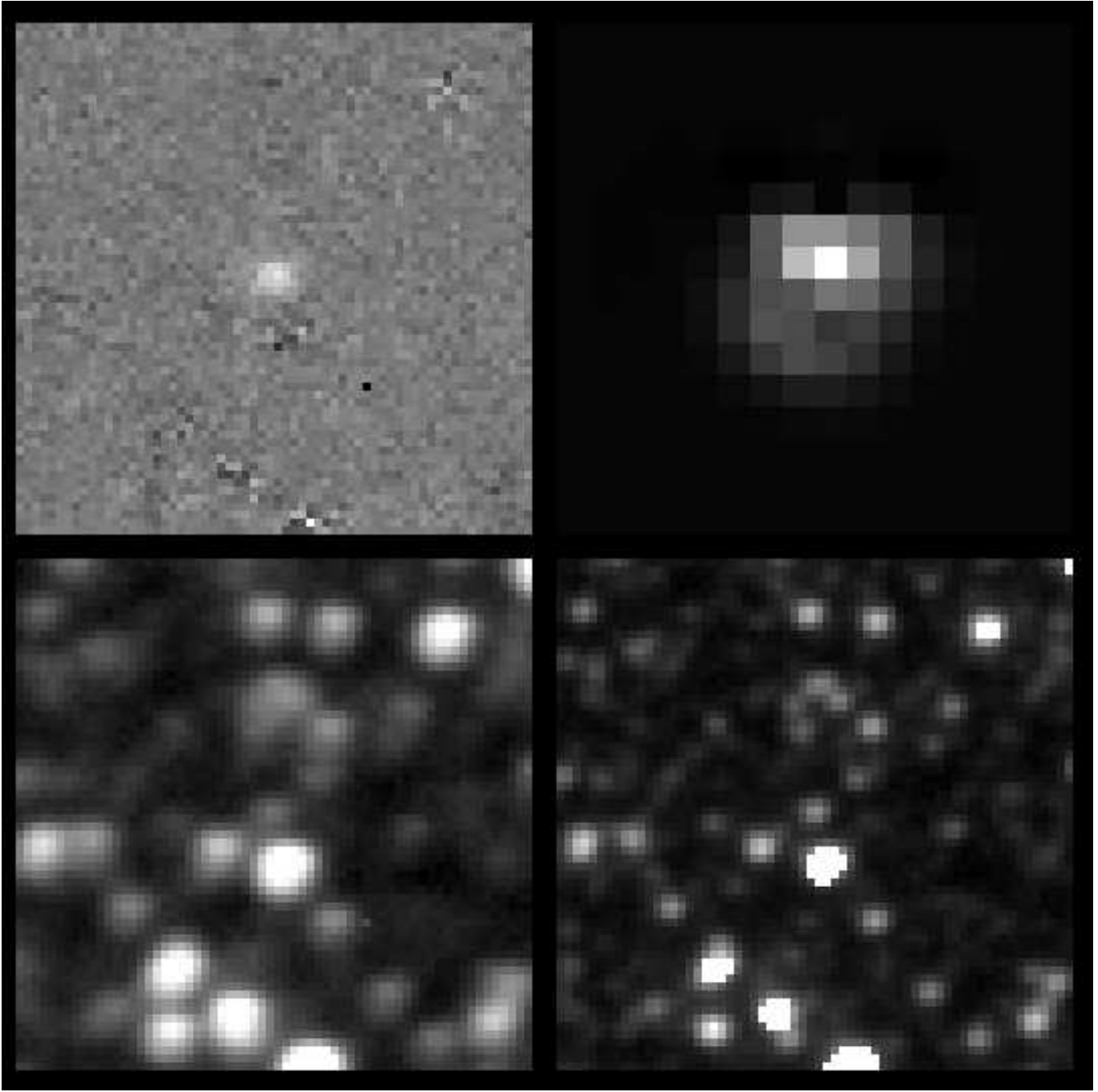,width=18cm}}
\caption{Example of subtracted image. The two bottom figures of the panel 
 are the original images. On the right is the reference image, and on the left
 is the image to be fitted by kernel convolution. The two upper figures
 show the best kernel solution on the right and on the left
the subtracted image.
 Note the complicated shape of the kernel.}
\end{figure*}
\begin{figure*}
\centerline{\psfig{angle=-90,figure=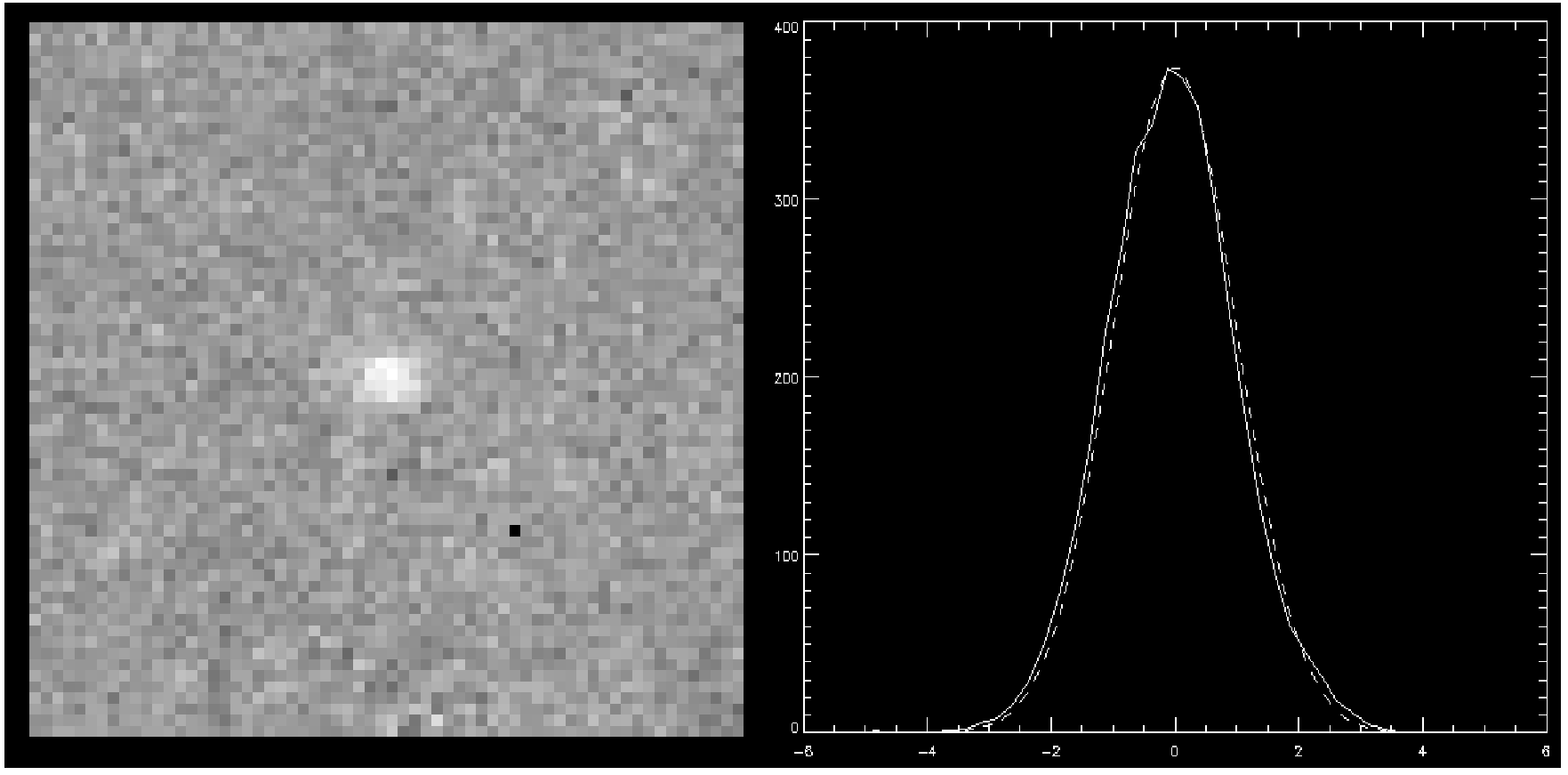,width=18cm}}
\caption{Noise in the subtracted image. The image on left is the subtracted
 image normalized by the Poisson deviations of both the reference and the
 image  for the small field presented at previous figure. On right we show
 the histogram of the pixels in this image. We sumperimposed to this histogram
 a gaussian of variance 1 (dashed line). Note that the deviations
 due to the bright stars are no longer visible. The variable star at center
 clearly stands out at very significant level. The dark pixel in the image is
 a cosmic.}
\end{figure*}
To spot the variables
stars, we created a ``deviation image'' by co-adding the square of the
subtracted images.
We normalized the deviation image by normalizing with the pixels standard
deviations. We found many variables at very significant levels. Most of
them seem to be bright giants with small amplitudes. Some of the
variables appeared to be periodic, we found a few RR Lyraes and some
eclipsing variables. We computed the flux variations for these stars
by making simple aperture photometry. In Fig. 3 and Fig. 4 we give an illustration of
the result we have obtained. 
\begin{figure*}
\centerline{\psfig{angle=-90,figure=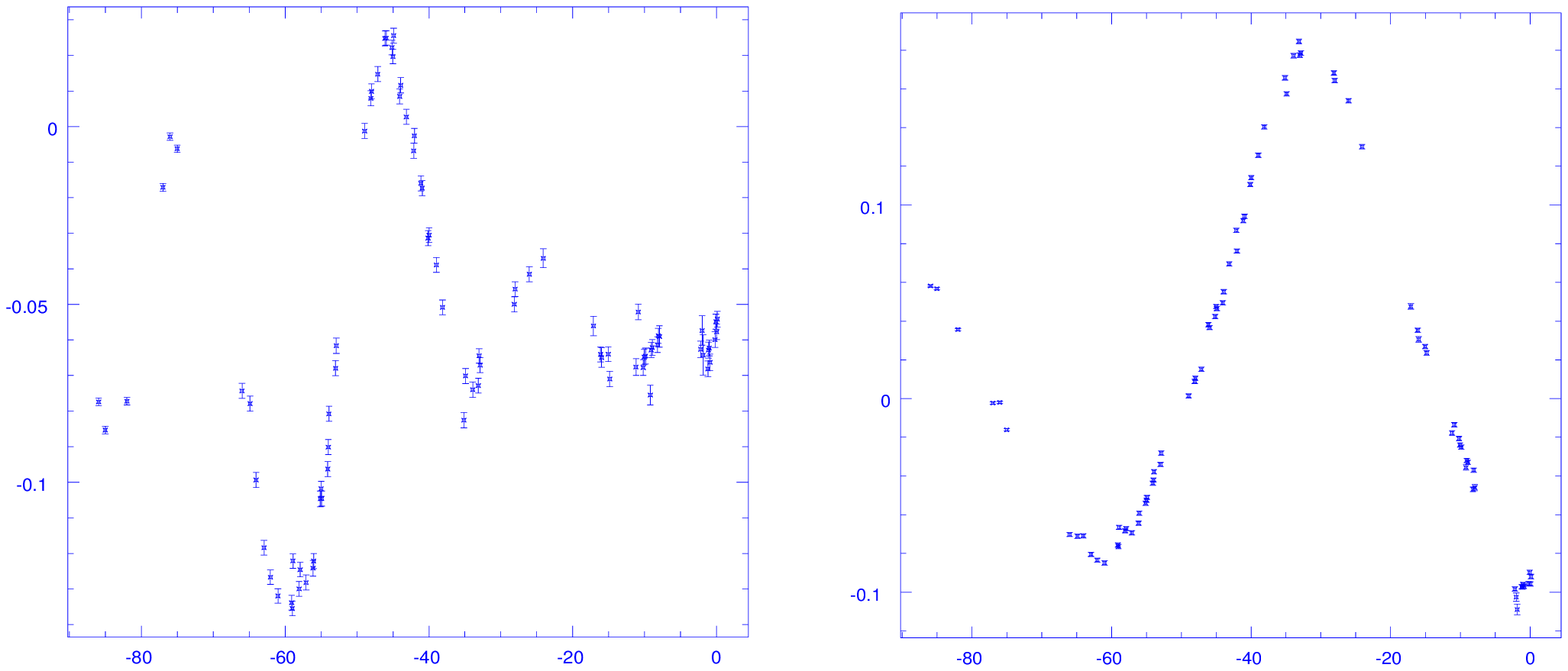,width=18cm}}
\caption{Examples of bright giants light curve. The x-axis are
 days, the y-axis are percentage variation (with respect to reference
 frame). Errors bars are derived from the Poisson deviations associated to each image,
 we do not include here the deviations associated with the reference
 image.} 
\end{figure*}
\begin{figure*}
\centerline{\psfig{angle=0,figure=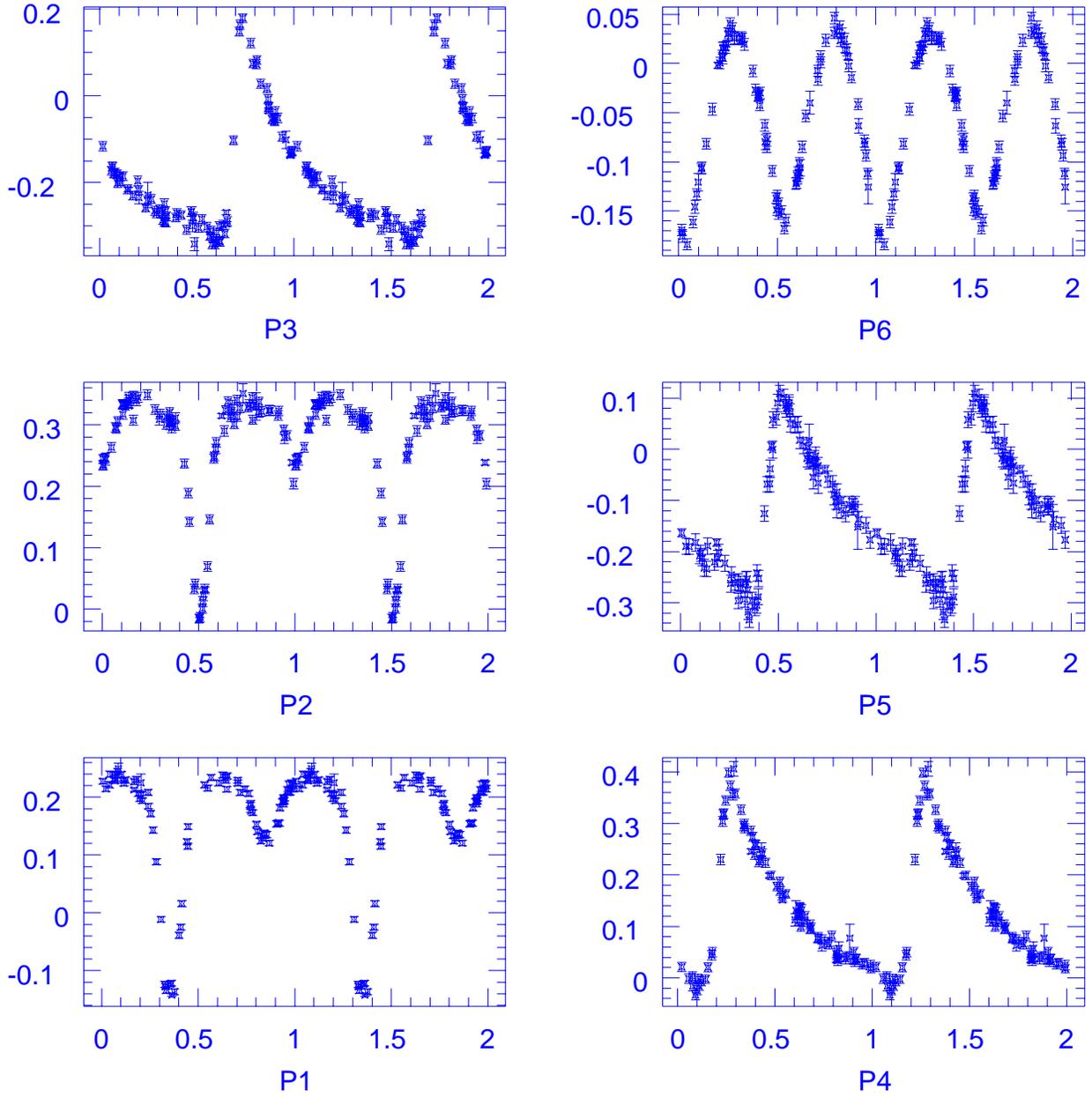,width=18cm}}
\caption{The periodic variables.The x-axis represent the phase
 , the y-axis are percentage variation (with respect to reference
 frame).}
\end{figure*}
\section{Sources of Noise in the Residual Image.}
As discussed above, the variance of the residual image is approximately
equal to the sum of the variances of the input images. If we created a
reference image by co-adding a large number of good-seeing images we could
remove the contribution of the noise in the reference; we would, of course,
have to be careful about variability between the different reference frames.
Upon inspection of the residual images, however, some showed significantly
larger residuals than expected from Poisson statistics near the position
of bright, but non-saturated, stars (it does represent less than 1 \% of the
 stars visible on the frame). These showed the characteristic
signature of centering errors, with equal positive and negative residuals
even for stars which show no evidence of variability (i.e. the sum of all
residuals within a few arc-seconds is zero). We believe that this is produced
by the turbulent atmosphere modulating our Kernel on the scale of our
sub-regions.

Shao and Colavita (1992) quote the
variance in the angle between two stars separated by $\theta$ as
$$
\sigma_\delta^2 \approx
        5.25 \left(\theta/\hbox{radian}\right)^{2/3}
        \left(t/\hbox{sec}\right)^{-1}
        \int C_n^2(h) h^{2/3} V^{-1}\!(h) \,dh
$$
for the regime in which we are interested (their equation 2). They evaluate
the integral using data from Roddier et al. (1990) for a night on Mauna
Kea with $\approx 0.5$arcsec seeing to give
$$
\sigma_\delta \approx
        1.1 \left(\theta/\hbox{radian}\right)^{1/3}
        \left(t/\hbox{sec}\right)^{-1/2} \hbox{arcsec}
$$
If we assume that the integral over the atmosphere scales with seeing
in a similar way to the integral
$$
        \int C_n^2(h) \,dh
$$
which enters into the definition of the Fried parameter, $r_0$, we may
expect that this result will scale as $(r_0 \lambda^{-6/5})^{-5/6}$
(a result which is independent of $\lambda$ due to the wavelength dependence
of $r_0$). In 1 arcsec seeing, therefore, we may expect that
$$
\sigma_\delta \approx
        2 \left(\theta/\hbox{radians}\right)^{1/3}
        \left(t/\hbox{sec}\right)^{-1/2} \hbox{arcsec}
$$
On the typical scale of our $128\times256$ regions, and for 128s
exposures, this corresponds to an RMS image motion of about 0.011arcsec.
If we model the PSF as a Gaussian with width parameter $\alpha$
($\alpha \approx 0.424$ for 1arcsec FWHM images), this would produce
a maximum residual of $\sigma_\delta/\alpha \exp(-1/2)$, or 1.6\%. This is of
the same order as the residuals that we see in our frames.
\section{Harmonic fitting to the periodic variables.}
We expect that the periodic variables light curves to be well
approximated with truncated Fourier series. We calculate the period
using the Renson method (1978), and we fit Fourier series with
different number of harmonics. The errors are calculated from the photon
noise in each image. We do not include the noise
associated with the reference image because it is produces an error only in
the total magnitude, and, to first order, doesn't affect the variable part
of the object's flux.
 We estimate each time the chi-square
per degree of freedom ($\chi^2_d$), and we look for the best chi-square
with the minimum number of harmonics.
The results are given in table
1 where we see that the resulting value of $\chi_d$ is close to unity,
for most variables. Except for variable P1 our mean error is at most
 only 25 $\%$ larger than the Poisson expectation (i.e. $\chi^2_d < 1.56$);
of course, this $\chi^2_d$ excess is significant.
In the case of the variable P1 the $\chi^2_d$ is very
 inconsistent with the Poisson expectation. This variable has about the same
 brightness as P6. We checked the quality of the subtracted images, but 
 could not identify any defects. The quality of the image subtraction is as
 good for P1 and for P6, they have about the same brightness, so what's wrong ?
 Considering that the mean error is fairly small (about 1\%), we might
suspect some residual error due to flat fielding. However, we get a mean
 residual of only 0.6 \% for P6 and $\chi^2_d=1.1$ showing that the flat
 fielding errors are much smaller than 0.6 \%. This is not surprising
 because these images were taken in drift scan mode, and consequently, we
 average the
 sensitivity of many pixels. We conclude that there must be some intrinsic 
 reason for P1's bad $\chi^2_d$. It is possible that variables do not repeat
 perfectly
 from cycle to cycle. This kind of variable star is well known to have spots
 which are likely to induce variability at the sub percent level. It is also
 possible that the RR Lyraes don't repeat perfectly, they are well known
 to show the Blashko effect, and we can explain some of the $\chi^2_d$ as
being due to cycle to cycle variations.
Although estimating the $\chi^2_d$ of periodic
variable stars is not an absolute test, we conclude that on average we are
 only about 20 \% above the Poisson error, and consequently there is not much
to 
be gained from improving our method. However, we must note
 that the errors due to the reference frame are the same for the integrated
 flux of a star on each image only at first order of approximation. By 
 convolving the reference each time to fit the seeing variations, we change 
slightly the
 noise distribution around the star. Especially for the case where a bright
 star is close to our object, convolving with the kernel might spread some
 noise into our photometric aperture. This effect will be negligible
 for good seeing frames, but noticeable when the seeing's bad. An obvious 
solution is to construct a reference with a signal to noise as good as 
possible by stacking the best seeing images; see the next section.
Another approach with potential to improve the signal-to-noise would be
to use a matched filter to measuring our stars variability. Unfortunately,
simply applying the usual PSF-filter leads to problems with aperture
corrections, and we shall not investigate this approach in this paper.
\begin{table*}[htb]
 \caption{Harmonics fitting to the periodic variables. In column 3
 we give the value of the mean residual to the fit. It is useful to 
 compare this residual to possible flat fielding errors. We also give an
 estimate of the star magnitude difference to the RR Lyrae $\Delta_{mag}$. We
 assume that the RR Lyrae have all the same mean magnitude. A crude estimate for the
 RR Lyrae mean magnitude in this field is ${\rm I} \simeq 17$.}
  \begin{flushleft}
  \begin{tabular}{llll}
   \hline\noalign{\smallskip}
   Variable & $\chi^2_d$ & Mean Residual (\%) & $\Delta_{mag}$ \\
   \hline\noalign{\smallskip}        
       P1 & 2.01 & 1.0 & -1.117\\
       P2 & 1.43 & 1.1 & -0.4402\\  
       P3 & 1.55 & 1.6 & 0\\  
       P4 & 1.46 & 1.2 & 0\\  
       P5 & 1.17 & 1.3 & 0\\  
       P6 & 1.1  & 0.6 & -0.4348\\  
    \noalign{\smallskip}
   \hline      
   \end{tabular}
  \end{flushleft}
\end{table*}
\section{Improving the Reference Frame.}
 We averaged the 20 best seeing images to build a reference frame with
 excellent signal to noise. The resulting seeing is of course not as good as
 it was in our previous reference which was the best image. But the seeing 
 variations are much reduced, as well as the noise amplitude. All the images
 were reprocessed using this new reference. We found that all the subtracted
 frames were improved. Even for the good seeing frame, were the seeing quality
 of the reference is critical we found some improvements. The light curves of 
 the variables stars were also improved, we give the result of harmonic fitting
 in table 2.
\begin{table*}[htb]
 \caption{Harmonics fitting to the light curves obtained with the new 
 reference. See table 1 for the meaning of columns.}
  \begin{flushleft}
  \begin{tabular}{lll}
   \hline\noalign{\smallskip}
   Variable & $\chi^2_d$ & Mean Residual (\%)\\ 
   \hline\noalign{\smallskip}        
       P1 & 2.0  & 1.0\\
       P2 & 1.16 & 1.0\\  
       P3 & 1.27 & 1.45 \\  
       P4 & 1.45 & 1.2 \\  
       P5 & 1.15 & 1.3\\  
       P6 & 1.03 & 0.6\\  
    \noalign{\smallskip}
   \hline      
   \end{tabular}
  \end{flushleft}
\end{table*}
\section{Computing time.}
 One might think that a method which fits all the pixels in an image
 (even if the fit is linear) is going to be much more time consuming
 than conventional methods. But the actual cost of the calculations 
 is much lighter than might appear at first glance.
Most of the computing time is taken by the calculation
 of the matrix we define in section 2.3 this is an $N^2$ process 
 (N is the number of basis function we use). The rest of the calculation
 is an N process. The matrix could be calculated once for all and used
 to fit the kernel solution for all images. A problem with this approach
is that we reject different pixels on each frame 
 (due to new saturated pixels, or variable stars) and consequently the matrix
 elements change.  In practice, we find that we reject no more than 1 \% percent
 of the total number of pixels,  so that all that we have to do is to calculate
 the matrix elements for the rejected pixels, and subtract them from the
 original values. This process cost very little CPU, and once the
 original matrix has been built, the kernel solution can be fitted very quickly
even though we use several clipping passes. The rest of the operations requires about
 the same computing time. By applying this method we can process a 
$1024 \times 1024$ frame in about 1 min with a 200 Mhz PC; this 
could certainly be improved further by using better numerical algorithms for
 the solution of the linear system.
\section{Acknowledgments}
 We would like to thank the OGLE team for providing the CCD images we 
 presented in our article. In particular we would like to thank A. 
 Udalski and M. Szyma\'nski for helping with the data. We are especially
 indebted to B. Paczy\'nski for supporting our project, and for many 
 interesting discussions. C. Alard would like to acknowledge support 
 from NSF grant AST-9530478 during his stay in Princeton, 
 where most of the research was done.
\end{document}